\documentclass[prd,aps,eqsecnum,nofootinbib]{revtex4}
\usepackage{epsfig,graphicx,latexsym}
\def\rmd{{\rm d}}

\def\MNRAS{{\em Mon. Not. Roy. Astron. Soc.~}}
\def\CQG{{\em Class. Quantum Grav.~}}
\def\PRD{{\em Phys. Rev. D~}}
\def\PR{{\em Phys. Rev.~}}

\newcommand{\erf}[1]{(\ref{#1})}
\newcommand{\fl}{\hspace*{-6pc}}

\begin{document}

\title{Electromagnetic Fields of Separable Space-Times}
\author{Jonathan R Gair\footnote{email address: jgair@ast.cam.ac.uk} and 
Donald
Lynden-Bell}
\affiliation{Institute of Astronomy, University of Cambridge,
Madingley Road, Cambridge, CB3 0HA, UK}

\date{\today}

\begin{abstract}
Carter derived the forms of the metric and the vector potentials of the space-times in which the relativistic Schr\"{o}dinger equation for the motion of a charged particle separates. Here we show that on each `spheroidal' surface a rotation rate, $\omega$, exists such that relative to those rotating axes the electric and magnetic fields are parallel and orthogonal to the spheroid which is thus an equipotential in those axes. 
\newline\indent All the finite Carter separable systems without magnetic monopoles or gravomagnetic NUT monopoles have the same gyromagnetic ratio as the Dirac electron.
\end{abstract}

\maketitle

\section{Introduction}
\label{intro}
It is well known that for any electromagnetic field in flat space, there exists a local Lorentz frame in which the measured electric and magnetic fields, ${\bf E'}$ and ${\bf B'}$, are parallel. This local Lorentz frame has velocity ${\bf v}/(1+v^2/c^2) = c({\bf E}\wedge{\bf B})/(E^2 + B^2)$ \cite{ctf}. In investigations of the `Magic' electromagnetic field given by ${\bf E} + {\rm i}{\bf B} = - \nabla \Phi$ where $\Phi = q/\sqrt{({\bf r} - {\rm i}{\bf a})^2}$, with ${\bf a} = (0,0,a)$, one of us discovered \cite{gairthesis} that the velocity of this special Lorentz frame for that field was
\begin{eqnarray}
&{\bf v}& = \omega \hat{\bf z} \wedge {\bf R} \nonumber \\
{\rm where} \qquad &\omega& = \frac{a\,c}{\tilde{r}^2 + a^2}  \qquad \mbox{and} \qquad {\bf R} = (x, y, 0)
\label{magicom}
\end{eqnarray}
and $\tilde{r}$ is the spheroidal coordinate, constant on the spheroid confocal with the equatorial disc of radius $a$, and which takes the value $z$ on that axis. Furthermore, at each point ${\bf E'}$ and ${\bf B'}$ are both parallel to $\nabla \tilde{r}$. Notice that the set of these locally defined velocities produce a velocity field which rotates uniformly on each spheroidal surface with angular velocity $\omega(\tilde{r})$. This suggests that the electric and magnetic fields measured on the surface $\tilde{r}=const.$ in a frame uniformly rotating with angular velocity $\omega(\tilde{r})$ would also be parallel to $\nabla \tilde{r}$.  We later found that this pretty theorem could be generalised to all the Carter separable electromagnetic fields \cite{dlb00} in flat space. We subsequently realised that this property was known for the Kerr-Newman metric, from which the `Magic' field is derived, and has proven invaluable for studying electromagnetic fields outside rotating black holes \cite{znaj77}.

The `Magic' field is the electromagnetic field left behind if we take the charged Kerr-Newman metric and set Newton's $G$ to zero. The space-time then becomes Minkowskian but the charge distribution and its rotation remain. The simplicity of the resulting electromagnetic field has led to several investigations by Newman \cite{newman02,newman05} and Lynden-Bell \cite{dlb03,dlb04a,dlb04b}.

The Kerr-Newman metric is just one member of a family of metrics for which the Hamilton-Jacobi and Schr\"{o}dinger equations both separate, as demonstrated by Carter \cite{carter68}. Here we show that the above theorem is true for all of these Schr\"{o}dinger separable space-times. More specifically, we show that for each `spheroidal' surface $\tilde{r}=const.$ there exists a uniform rotation rate $\omega$ such that the electric field ${\bf E'}$ and magnetic field ${\bf B'}$ seen on that surface in the rotating axes are normal to that spheroid. In the following, we use units such that $c=1$ for convenience. For consistency with Carter, we will use the greek letters $\lambda$ and $\mu$ to denote two spatial coordinates in the spacetimes. We will also use greek letters for spacetime indices. However, we will use only $\alpha$, $\beta$, $\gamma$ and $\delta$ for dummy spacetime indices that vary over all coordinates. Whenever a tensor appears with an index $\lambda$ or $\mu$ it will denote that particular spatial component of the tensor.

\section{A demonstration of this property}
\label{proof}
\subsection{Carter separable spacetimes}
With a suitable relabelling of his coordinates Carter showed that the metrics and electromagnetic fields in which Schr\"{o}dinger's equation separates take the forms\footnote{Carter actually writes these expressions in terms of two arbitrary Killing vectors $\psi$ and $\chi$ but we have replaced the former with $t$ and regard this as a time coordinate. We regard $\chi$ as an azimuthal coordinate.}
\begin{equation}
\fl\rmd s^2 =g_{\alpha\beta}\rmd x^{\alpha} x^{\beta} = \frac{Z}{\Delta_{\lambda} }\rmd\lambda^2 +  \frac{Z}{\Delta_{\mu}} \rmd\mu^2 + \frac{\Delta_{\mu}}{Z}\,\left(P_{\lambda}\,\rmd t - Q_{\lambda}\,\rmd \chi\right)^2 - \frac{\Delta_{\lambda}}{Z}\,\left(P_{\mu}\,\rmd t - Q_{\mu}\,\rmd \chi\right)^2 \label{cartermet}
\end{equation}
and the simplest associated electromagnetic potential $A_{\gamma}$ is
\begin{equation}
\fl A_{\gamma}\,\rmd x^{\gamma} = \left(\frac{P_{\lambda}\,X_{\mu} + P_{\mu}\,X_{\lambda}}{Z}\right)\,\rmd t - \left(\frac{Q_{\lambda}\,X_{\mu} + Q_{\mu}\,X_{\lambda}}{Z}\right) \,\rmd \chi \label{carterEMF}
\end{equation}
where $\Delta_{\lambda}$, $P_{\lambda}$, $Q_{\lambda}$, $X_{\lambda}$ are functions of the spheroidal coordinate $\lambda$ alone and $\Delta_{\mu}$, $P_{\mu}$, $Q_{\mu}$, $X_{\mu}$ are functions of the other meridional coordinate $\mu$ alone. Finally
\begin{equation}
Z = P_{\lambda}\,Q_{\mu} - P_{\mu}\,Q_{\lambda}
\label{Zdef}
\end{equation}
and there is a restriction on the $P$'s and $Q$'s that $Z$ must be the sum of a function of $\lambda$ only and a function of $\mu$ only.

If we choose the functions to be
\begin{eqnarray}
\fl\Delta_{\lambda} &=& \lambda^2 + a^2 + G\,e^2-2GM\,\lambda, \qquad P_{\lambda} = -1, \qquad Q_{\lambda}= -(\lambda^2 + a^2), \qquad X_{\lambda}=e\lambda \nonumber \\
\fl\Delta_{\mu} &=& a^2-\mu^2, \qquad P_{\mu} = 1, \qquad Q_{\mu} = a^2-\mu^2, \qquad X_{\mu} = 0 \label{KNfncs}
\end{eqnarray}
and make a change of coordinates by setting $\lambda = r$, $\mu = a\cos\theta$ and $\chi=\phi/a$ then the metric and electromagnetic field reduce to the Kerr-Newman solution written in Boyer-Lindquist coordinates
\begin{eqnarray}
\rmd s^2 &=&- \left(\frac{\Delta - a^2\sin^2\theta}{\Sigma}\right) \rmd t^2 - \frac{2a\sin^2\theta(r^2+a^2-\Delta)}{\Sigma} \rmd t \rmd \phi  +\frac{(r^2+a^2)^2-\Delta a^2\sin^2\theta}{\Sigma} \sin^2\theta\rmd \phi^2 +\frac{\Sigma}{\Delta} \rmd r^2 + \Sigma\rmd\theta^2 \nonumber \\
A_{\gamma}\rmd x^{\gamma} &=& -\frac{e\,r}{\Sigma} \rmd t + \frac{e\,a\,r\,\sin^2\theta}{\Sigma}\,\rmd\phi \nonumber \\
{\rm where} & & \Delta = r^2+a^2+G\,e^2-2GM\,r, \qquad \Sigma = r^2+a^2\cos^2\theta \label{KNmet}.
\end{eqnarray}
The $G\rightarrow0$ limit of this equation reduces to flat space written in oblate spheroidal coordinates and the `Magic' field.

\subsection{Electromagnetic field measured by a static observer}
An observer in the metric \erf{cartermet} with constant $\lambda$, $\mu$ and $\chi$ coordinates moves on the worldline $u^{\alpha} = (1/\sqrt{-g_{tt}},0,0,0)$. An orthonormal tetrad based on this wordline, with legs oriented in the $\lambda$ and $\mu$ coordinate directions is given by
\begin{eqnarray}
\fl {\bf e}_{0}^{\alpha} &=& (1/\sqrt{-g_{tt}},0,0,0), \qquad {\bf e}_{1}^{\alpha} = (0,1/\sqrt{g_{\lambda\lambda}},0,0), \qquad {\bf e}_{2}^{\alpha} = (0,0,1/\sqrt{g_{\mu\mu}},0), \nonumber \\ \fl {\bf e}_{3}^{\alpha} &=& \left(\frac{g_{t\chi}}{\sqrt{-g_{tt}(g_{t\chi}^2-g_{tt}g_{\chi\chi})}},0,0,\sqrt{\frac{-g_{tt}}{g_{t\chi}^2-g_{tt}g_{\chi\chi}}}\right)
\label{stattet}
\end{eqnarray}
where we have taken the coordinates in the order $(t,\lambda,\mu,\chi)$. The electromagnetic field tensor measured in the tetrad frame is given by
\begin{equation}
F_{ij} = {\bf e}_{i}^{\alpha} \,  {\bf e}_{i}^{\beta} \, F_{\alpha\beta} =  {\bf e}_{i}^{\alpha} \,  {\bf e}_{i}^{\beta} \, \left(\partial_{\alpha}A_{\beta} - \partial_{\beta}A_{\alpha}\right) 
\end{equation}
where $A_{\alpha}$ is the electromagnetic potential given in \erf{carterEMF}. The measured electric field, given by $E_i=F_{0i}$, is
\begin{equation}
{\bf E} = \left(\frac{F_{0\lambda}}{\sqrt{-g_{tt}g_{\lambda\lambda}}}, \frac{F_{0\mu}}{\sqrt{-g_{tt}g_{\mu\mu}}}, \frac{F_{0\chi}}{\sqrt{g_{t\chi}^2-g_{tt}g_{\chi\chi}}} \right) \label{Edef}
\end{equation}
The measured magnetic field, given by $F_{ij} = -\epsilon_{ijk} B_{k}$, is
\begin{eqnarray}
\fl{\bf B} &=& \left(  \frac{g_{tt}F_{\mu\chi}-g_{t\chi} F_{\mu0}}{\sqrt{-g_{tt}g_{\mu\mu}(g_{t\chi}^2-g_{tt}g_{\chi\chi})}}, \frac{g_{t\chi} F_{\lambda0}-g_{tt}F_{\lambda\chi}}{\sqrt{-g_{tt}g_{\lambda\lambda}(g_{t\chi}^2-g_{tt}g_{\chi\chi})}}, -\frac{1}{\sqrt{g_{\lambda\lambda} g_{\mu\mu}}} F_{\lambda\mu} \right) \nonumber \\
\fl &=&  \left(- \frac{\sqrt{g_{\mu\mu} (g_{t\chi}^2 - g_{tt}g_{\chi\chi})}}{\sqrt{-g_{tt}}} \,F^{\mu\chi}, \frac{\sqrt{g_{\lambda\lambda} (g_{t\chi}^2 - g_{tt}g_{\chi\chi})}}{\sqrt{-g_{tt}}} \,F^{\lambda\chi}, - \sqrt{g_{\lambda\lambda} g_{\mu\mu}}F^{\lambda\mu} \right) \label{Bdef}
\end{eqnarray}
From these expressions we see that if we can find a rotating frame in which $F_{0'\mu'} = F_{0'\chi'} = F^{\lambda'\chi'} = F^{\lambda'\mu'} = 0$ on a spheroidal surface $\lambda' = const.$, then the measured electric and magnetic fields will be parallel, and in the direction of $\nabla \lambda'$, on that spheroid. This is the requirement of our theorem.

We note that the observer discussed here is not necessarily physical in general, since a $\chi = const.$ surface will not be timelike inside the stationary limit surface $g_{tt}=0$. However, we will show below that such observers are always physical in the special rotating frame that satisfies our theorem.

\subsection{Electromagnetic field in a uniformly rotating frame}
We now consider transforming \erf{cartermet} to a frame rotating with angular velocity $\omega$, by writing $\chi' = \chi - \omega t$, $t'=t$, $\lambda'=\lambda$, $\mu'=\mu$. For simplicity, we are defining rotation with respect to the $\chi$, $t$ frame, i.e., a ``rotating frame'' is one with non-zero angular velocity measured in $\chi$, $t$ coordinates. For asymptotically flat spacetimes of this type, then there is a notion of global rotation and we can assume that the $\chi$ Killing vector has been chosen to be the one that is non-rotating with respect to infinity. We will see in equation \erf{critom} below that the angular velocities of interest depend on $\lambda$ and so even if we do not make this special choice for $\chi$ the majority of these frames are rotating under either definition.

Under the transformation to the $\chi',t'$ coordinates defined above the electromagnetic potential and metric transform as
\begin{eqnarray}
\fl A_{t'} &=& A_t+\omega A_{\chi} = \frac{(P_{\lambda} -\omega Q_{\lambda})X_{\mu} + (P_{\mu} -\omega Q_{\mu})X_{\lambda}}{Z}, \nonumber \\ \fl A_{\chi'}&=&A_{\chi}= \left(\frac{Q_{\lambda}\,X_{\mu} + Q_{\mu}\,X_{\lambda}}{Z}\right) \nonumber \\
\fl g_{t't'} &=& g_{tt}+2\omega g_{t\chi} + \omega^2 g_{\chi\chi}, \qquad g_{t'\chi'}=g_{t\chi}+\omega g_{\chi\chi}, \qquad g_{\chi'\chi'}=g_{\chi\chi} \nonumber \\
\fl g^{t't'} &=& \frac{g_{\chi\chi}}{g_{tt}g_{\chi\chi} - g_{t\chi}^2}, \qquad g^{t'\chi'} = -\frac{g_{t\chi}+\omega g_{\chi\chi}}{g_{tt}g_{\chi\chi} - g_{t\chi}^2}, \qquad g^{\chi'\chi'} = \frac{g_{tt}+2\omega g_{t\chi} + \omega^2 g_{\chi\chi}}{g_{tt}g_{\chi\chi} - g_{t\chi}^2} .
\end{eqnarray}
We note that this transformation is equivalent to setting $P_{\lambda} \rightarrow (P_{\lambda}-\omega Q_{\lambda})$ and $P_{\mu} \rightarrow (P_{\mu}-\omega Q_{\mu})$ in Carter's general expressions \erf{cartermet}--\erf{carterEMF}. The $\mu$ component of the electric field, $F_{0'\mu'}$, becomes
\begin{equation}
F_{0'\mu'} = \partial_{t'} A_{\mu'} -\partial_{\mu'} A_{t'} = -\partial_{\mu'} A_{t'} = \frac{1}{Z} \left(P_{\lambda} - \omega Q_{\lambda}\right) \left(\frac{\rmd X_{\mu}}{\rmd \mu} + \frac{X_{\lambda}}{Z} \left(Q_{\mu} \frac{\rmd P_{\mu}}{\rmd \mu} - P_{\mu} \frac{\rmd Q_{\mu}}{\rmd \mu} \right) \right)
\end{equation}
It is clear that if we choose
\begin{equation}
\omega = \omega_c = \frac{P_{\lambda} ( \lambda_c)}{Q_{\lambda} (\lambda_c)}
\label{critom}
\end{equation}
then $F_{0'\mu'}=0$ everywhere on the spheroid $\lambda = \lambda' = \lambda_c$. We now consider the $\mu'$ component of the magnetic field, which is given by $F^{\lambda'\chi'}$ as we saw above.
\begin{eqnarray}
\fl F^{\lambda'\chi'} &=& g^{\lambda'\alpha'}g^{\phi'\beta'}\left(\partial_{\alpha'}A_{\beta'} - \partial_{\beta'}A_{\alpha'}\right)  = \frac{1}{g_{\lambda'\lambda'}} \left(g^{\chi'\chi'} \partial_{\lambda'} A_{\chi'} + g^{t'\chi'} \partial_{\lambda'} A_{t'} \right) \nonumber \\
\fl &=& \frac{1}{g_{\lambda\lambda}(g_{tt}g_{\chi\chi} - g_{t\chi}^2)} \left((g_{tt} + \omega g_{t\chi})\partial_{\lambda} A_{\chi} - (g_{t\chi} + \omega g_{\chi\chi}) \partial_{\lambda}A_{t} \right) \nonumber \\
\fl &=&  \frac{(P_{\lambda}-\omega Q_{\lambda})}{g_{\lambda\lambda}(g_{tt}g_{\chi\chi} - g_{t\chi}^2) \,Q_{\lambda}Z}  \left((\Delta_{\lambda}P_{\mu}Q_{\mu} - \Delta_{\mu}P_{\lambda}Q_{\lambda}) \frac{\partial}{\partial \lambda}\left(\frac{Q_{\lambda}X_{\mu}+Q_{\mu}X_{\lambda}}{Z} \right) \right. \nonumber \\
\fl && \hspace{1.6in} + \left. (\Delta_{\mu} Q_{\lambda}^2 - \Delta_{\lambda}Q_{\mu}^2) \frac{\partial}{\partial \lambda}\left(\frac{P_{\lambda}X_{\mu}+P_{\mu}X_{\lambda}}{Z} \right) \right) + \tilde{F}^{\lambda'\chi'} \label{specB}
\end{eqnarray}
In \erf{specB} we have grouped some additional terms into the expression $\tilde{F}^{\lambda'\chi'}$. These may be shown to vanish
\begin{eqnarray}
\fl \tilde{F}^{\lambda'\chi'} &\propto& \left[ \Delta_\lambda Q_\lambda P_\mu Q_\mu^2 - P_\lambda \Delta_\lambda Q_\mu^2 P_\mu -\Delta_\lambda Q_\lambda P_\mu Q_\mu^2 + P_\lambda \Delta_\lambda Q_\mu^2 P_\mu \right]  \frac{\partial}{\partial \lambda}\left(\frac{X_{\lambda}}{Z}\right) \nonumber \\ \fl &+& X_{\mu} \left[(\Delta_\lambda P_\mu^2 Q_\lambda - P_\lambda \Delta_\lambda Q_{\mu} P_\mu) \frac{\partial}{\partial \lambda}\left(\frac{Q_{\lambda}}{Z}\right)  + (\Delta_\lambda P_\lambda Q_\mu^2 - Q_\lambda \Delta_\lambda Q_{\mu} P_\mu) \frac{\partial}{\partial \lambda}\left(\frac{P_{\lambda}}{Z}\right) \right] \nonumber \\ \fl &=& 0 \nonumber  
\end{eqnarray}
The first square-bracketed term manifestly vanishes, and the vanishing of the second term follows from using the definition of $Z$ in equation \erf{Zdef}. From equation \erf{specB} it is clear that choosing the special $\omega$ \erf{critom} causes the $\mu'$ component of the magnetic field to vanish also. The $\chi'$ components of the electric ($F_{0'\chi'}$) and magnetic ($F^{\lambda'\mu'}$) fields can be seen to vanish trivially. In this special rotating frame, therefore, the electric and magnetic fields are parallel and in the direction of $\nabla \lambda'$ everywhere on the spheroid $\lambda'=\lambda_c$. Since $\lambda_c$ is arbitrary, we can find such a frame for every spheroidal surface in the space-time. This is our theorem. The angular velocity \erf{critom} is the angular velocity of the Carter tetrad \cite{carter73}, i.e., an observer at rest in the Carter tetrad frame will have angular velocity \erf{critom} as measured in the $\chi$, $t$ frame. Indeed, if we define the tetrad \erf{stattet} for an observer at rest in the $\chi'$, $t'$ frame that is rotating at the critical angular velocity $\omega_c(\lambda_c)$, then it coincides with the Carter tetrad on the spheroid $\lambda' = \lambda_c$.

For the special case of the Kerr-Newman field, the angular velocity of the special frame is
\begin{equation}
\frac{\rmd \phi}{\rmd t} = a \frac{\rmd \chi}{\rmd t} = a \omega = \frac{a}{r^2+a^2}
\end{equation}
as previously given in equation \erf{magicom} and discussed in \cite{znaj77}. We note further that in this rotating frame the $g_{t't'}$ component of the metric is given by
\begin{equation}
g_{t't'} = -\frac{\Delta_{\lambda}}{Z} (P_{\mu} - \omega Q_{\mu})^2 + \frac{\Delta_{\mu}}{Z} (P_{\lambda} - \omega Q_{\lambda})^2
\end{equation}
In the vicinity of the spheroid $\lambda'= \lambda_c$ it is clear that $g_{t't'} < 0$, so an observer with constant $\chi'$, $\lambda'$ and $\mu'$ coordinates is timelike and our earlier conclusions about how to interpret the components of $F_{\alpha\beta}$ is valid. We note also that
\begin{equation}
g_{t'\chi'}^2-g_{t't'}g_{\chi'\chi'} = g_{t\chi}^2-g_{tt}g_{\chi\chi}  = \Delta_{\lambda}\Delta_{\mu} > 0
\end{equation}
so all terms in \erf{Edef} and \erf{Bdef} are well defined.

The other components of the electromagnetic field tensor, $F_{0'\lambda'}$ and $F^{\mu'\chi'}$ may be found easily by noting that under the transformation $\lambda' \leftrightarrow \mu'$, the $(t', \chi')$ metric components are unchanged but the components of $A_{\gamma'}$ change sign. We find
\begin{eqnarray}
\fl F_{0'\lambda'} &=& \frac{\left(P_{\mu} - \omega Q_{\mu}\right)}{\left(Q_{\lambda}P_{\mu} - P_{\lambda} Q_{\mu} \right)} \left(\frac{\rmd X_{\lambda}}{\rmd \lambda} + \frac{X_{\mu}}{Z} \left(Q_{\lambda} \frac{\rmd P_{\lambda}}{\rmd \lambda} - P_{\lambda} \frac{\rmd Q_{\lambda}}{\rmd \lambda} \right) \right) \\
\fl F^{\mu'\chi'} &=& 
\frac{-(P_{\mu}-\omega Q_{\mu})\Delta_{\lambda}}{g_{\lambda\lambda}(g_{tt}g_{\chi\chi} - g_{t\chi}^2) \,Z} \left(\frac{\rmd X_{\mu}}{\rmd \mu} + \frac{1}{Z} \left((X_{\lambda}Q_{\mu} +X_{\mu}Q_{\lambda}) \frac{\rmd P_{\mu}}{\rmd \mu} - (X_{\lambda}P_{\mu} +X_{\mu}P_{\lambda}) \frac{\rmd Q_{\mu}}{\rmd \mu} \right)\right)
\end{eqnarray}
In the special frame $\omega=\omega_c$ and on the surface $\lambda'=\lambda_c$ these reduce to
\begin{eqnarray}
\fl F_{0'\lambda'} &=& \frac{1}{Q_{\lambda}} \left(\frac{\rmd X_{\lambda}}{\rmd \lambda} + \frac{X_{\mu}}{Z} \left(Q_{\lambda} \frac{\rmd P_{\lambda}}{\rmd \lambda} - P_{\lambda} \frac{\rmd Q_{\lambda}}{\rmd \lambda} \right) \right) \label{Elamcrit} \\
\fl F^{\mu'\chi'} &=& \frac{-\Delta_{\lambda}}{g_{\lambda\lambda}(g_{tt}g_{\chi\chi} - g_{t\chi}^2) \,Q_{\lambda}} \left(\frac{\rmd X_{\mu}}{\rmd \mu} + \frac{1}{Z} \left((X_{\lambda}Q_{\mu} +X_{\mu}Q_{\lambda}) \frac{\rmd P_{\mu}}{\rmd \mu} - (X_{\lambda}P_{\mu} +X_{\mu}P_{\lambda}) \frac{\rmd Q_{\mu}}{\rmd \mu} \right)\right) \label{Blamcrit}
\end{eqnarray}
where both \erf{Elamcrit} and \erf{Blamcrit} are to be evaluated at $\lambda = \lambda' = \lambda_c$.

We can also evaluate the mixed energy-momentum tensor of the electromagnetic field in the rotating frame, on the spheroid $\lambda'=\lambda_c$. We find
\begin{eqnarray}
\fl T^{\alpha'}\,_{\beta'} &=& \frac{1}{\mu_{0}} \left( F^{\alpha'\gamma'}F_{\gamma'\beta'} + \frac{1}{4} \delta^{\alpha'}_{\beta'} F^{\gamma'\delta'}F_{\gamma'\delta'} \right) = \frac{1}{\mu_0} \left( \begin{array}{cccc}W&0&0&2g_{t'\chi'}\, W /g_{t't'}\\ 0&W&0&0 \\ 0&0&-W&0 \\ 0&0&0&-W
\end{array} \right) \\
\fl {\rm where}\mbox{ } W &=& \frac{1}{2} \left(F_{\mu'\chi'}F^{\mu'\chi'} - F_{t'\lambda'}F^{t'\lambda'}\right)= \frac{1}{-g_{t't'}} \left( \frac{1}{g_{\lambda'\lambda'}}(F_{t'\lambda'})^2+ g_{\mu'\mu'}(g_{t'\chi'}^2-g_{t't'}g_{\chi'\chi'}) (F^{\mu'\chi'})^2\right) .
\end{eqnarray}
This is of diagonal form, except for the $T^{t'}\,_{\chi'}$ term which arises from the non-diagonality of the metric. It is easily seen that the energy-momentum tensor measured in the tetrad \erf{stattet} of the static observer is $T_{ij} = T_{\alpha'\beta'}{\bf e}^{\alpha'}_{(i)} {\bf e}^{\beta'}_{(j)} = {\rm diag}(-W,W,-W,-W)$. The term $W$ is the energy density of the electromagnetic field, $({\bf E}^2 + {\bf B}^2)/8\pi$, so this is of the usual form expected in the frame where ${\bf E}$ and ${\bf B}$ are parallel.

\subsection{A few remarks}
In the above, we have chosen to interpret $\lambda$ as the `radial' spheroidal coordinate, and $\mu$ as the meridional coordinate. Under the transformation $\lambda \leftrightarrow \mu$, $P \leftrightarrow Q$, the metric \erf{cartermet} is unchanged, except for the signs of the $g_{tt}$, $g_{t\chi}$ and $g_{\chi\chi}$ terms. This transformation therefore changes the interpretation of the two killing vectors. If we interpret $t$ as the timelike killing vector, then it is possible to find a rotating frame with velocity $\omega_c$ in which the electric and magnetic fields are parallel on a surface $\lambda = constant$. If we interpret $\chi$ as the timelike killing vector, then in a frame rotating with angular velocity $\rmd t/\rmd\chi = - Q_{\mu}/P_{\mu}$, the fields will be parallel on a surface $\mu = constant$.

We can also interpret $\chi$ not as an azimuthal killing vector but as a translational killing vector, i.e., a ``$z$'' coordinate. In that case, our theorem shows that it is possible to find a frame moving in the z direction with uniform velocity in which the electric and magnetic fields are parallel and perpendicular to the cylindrical surface $\lambda = constant$.

Our result implies that if we have a spheroidal shell of charge at $\lambda = \lambda_c$, rotating at angular velocity $\omega = \omega_c =  P_\lambda(\lambda_c)/Q_\lambda(\lambda_c)$, then the force experienced by the shell is purely `radial' in the sense of being parallel to $\nabla\lambda$ and normal to the spheroidal shell. This is most easily seen if we work in the frame rotating at angular velocity $\omega_c$. In that frame, the current distribution is just $j^{\beta'} = (\rho,0,0,0)$ and the force on the shell is
\begin{equation}
f^{\alpha'} = F^{\alpha'}_{\beta'} \, j^{\beta'} = \rho \, g^{\alpha'\sigma'} \, F_{\sigma' 0'} = \rho\,\delta^{\alpha'}_{\lambda'} \, F_{\lambda' 0'} .
\end{equation}
Where the latter follows from the fact that $F_{\lambda' 0'}$ is the only non-vanishing $F_{\alpha' 0'}$ in the rotating frame. Since the $\lambda$ direction does not change when we transform to the rotating frame, $f^\alpha$ is in the radial direction in the original coordinate system also. The force is thus in the direction of $\nabla \lambda$.

An interesting interpretation of the critical velocity $\omega$ arises from the fact that it is the angular velocity of the Carter tetrad. The independence of the metric on $t$ and $\chi$ provides two first integrals for geodesic motion, an ``energy'' $E=u_t$ and ``z-component of angular momentum'' $L_z=u_{\chi}$
\begin{eqnarray}
\Delta_{\mu} \left( P_\lambda \dot{t} - Q_\lambda \dot{\chi}\right) = Q_\mu E + P_\mu L_z \label{geo1} \\
\Delta_{\lambda} \left( P_\mu \dot{t} - Q_\mu \dot{\chi}\right) = Q_\lambda E + P_\lambda L_z \label{geo2} . 
\end{eqnarray}
Here a dot denotes differentiation with respect to an affine parameter. We notice that if $E/L_z = -P_\mu/Q_\mu$ initially, then $\rmd \chi/\rmd t = P_\lambda/Q_\lambda = \omega_c$. The angular velocity will maintain this value if $P_\mu/Q_\mu$ is constant. In general, this means that such a special geodesic must be in a $\mu = constant$ surface. The remaining two geodesic equations may be found from conservation of the length of the tangent vector to the geodesic, and from the Euler-Lagrange equation for motion in $\mu$
\begin{eqnarray}
\fl \mbox{\hspace{0.75in}}\epsilon &=& \frac{Z}{\Delta_\lambda} \dot{\lambda}^2 + \frac{Z}{\Delta_\mu} \dot{\mu}^2 + \frac{1}{\Delta_\mu \,Z} \left(Q_\mu E + P_\mu L_z\right)^2 - \frac{1}{\Delta_\lambda \,Z} \left(Q_\lambda E + P_\lambda L_z\right)^2 \label{geo3} \\
\fl \frac{\rmd}{\rmd \tau} \left(\frac{2Z}{\Delta_\mu}\,\dot{\mu} \right) &=& (P_\lambda Q_\mu' - Q_\lambda P_\mu') \epsilon + \frac{2\Delta_\lambda}{Z^2} (P_\mu Q_\mu'-Q_\mu P_\mu')(P_\mu \dot{t} - Q_\mu \dot{\chi}) (P_\lambda \dot{t} - Q_\lambda \dot{\chi}) \nonumber \\ && + \left(\frac{\Delta_\mu'}{Z} - \frac{2\Delta_\mu}{Z^2} (P_\lambda Q_\mu' - Q_\lambda P_\mu')\right) (P_\lambda \dot{t} - Q_\lambda \dot{\chi})^2 -\frac{Z\,\Delta_\mu'}{\Delta_\mu^2} \dot{\mu}^2 \label{geo4}
\end{eqnarray}
In this $\epsilon=-1$ for timelike geodesics and $\epsilon=0$ for null geodesics and we use a dash ($'$) to denote differentiation of a function with respect to its argument (in this case $\mu$). It is clear from the above that if we initially choose $\dot{\mu} = 0$ and $\dot{\chi}/\dot{t} = P_\lambda/Q_\lambda$, then all terms in \erf{geo4} vanish except for $\epsilon\,(P_\lambda Q_\mu'-Q_\lambda P_\mu')$. This is zero for null geodesics. This special family of null geodesics is the family of geodesics that define the Carter tetrad --- photons move with purely `radial' velocity in this tetrad frame (i.e., in the $\lambda$ direction). We conclude that another way to interpret the separability conditions of this class of spacetimes is that the electric and magnetic fields measured in the natural orthonormal frame of the spacetime must both be parallel to the null geodesics observed in that frame.

In reflection symmetric spacetimes (i.e., spacetimes for which $P_\mu$, $Q_\mu$, $\Delta_\mu$ and $X_\mu$ are even functions of $\mu$), timelike geodesics exist in the equatorial plane, $\mu = 0$. This is consistent with the above analysis, since for such spacetimes $Q_\mu' = P_\mu'=0$ when $\mu = 0$. Thus, there is also a family of timelike equatorial geodesics for which $\rmd\chi/\rmd t=\omega_c(\lambda)$. If we consider the motion of charged particles, then a particle moving with the critical angular velocity will experience a force purely in the $\lambda$ direction. In Newtonian theory, a particle moving under a central force has conserved angular momentum. We saw in equation \erf{geo1} that an orbit with the critical angular velocity corresponds to a constant angular momentum. This suggests that there might exist charged particle orbits that are traversed at $\omega = \omega_c$. This is indeed the case. Equation \erf{geo1} is true for any vector $u^\alpha$ is we replace $\dot{t} \rightarrow u^t$, $\dot{\chi} \rightarrow u^\chi$, $E \rightarrow u_t$ and $L_z \rightarrow u_\chi$. Under the action of an electromagnetic force the geodesic equation becomes $Du_\alpha/D\tau = (q/G m) F_{\alpha\beta} u^\beta$. Differentiation of equation \erf{geo1} with respect to the affine parameter gives
\begin{eqnarray}
\frac{\rmd}{\rmd \tau} \left(\Delta_\mu \left(P_\lambda u^t - Q_\lambda u^\chi \right)\right) &=& \frac{q}{G m} u^\lambda \left( P_\mu \partial_\lambda\left( \frac{Q_\lambda X_\mu + Q_\mu X_\lambda}{Z} \right) - Q_\mu \partial_\lambda\left( \frac{P_\lambda X_\mu + P_\mu X_\lambda}{Z} \right) \right) \nonumber \\ && \hspace{0in}+ \frac{q}{G m} u^\mu \left( P_\mu \partial_\mu\left( \frac{Q_\lambda X_\mu + Q_\mu X_\lambda}{Z} \right) - Q_\mu \partial_\mu\left( \frac{P_\lambda X_\mu + P_\mu X_\lambda}{Z} \right) \right)  + u^\mu \left(Q_\mu' u_t + P_\mu'u_\chi \right) \nonumber \\
&& \hspace{0in} = u^\mu \left( Q_\mu' u_t + P_\mu'u_\chi + \frac{q}{G m} \left( \frac{X_\lambda (P_\mu Q_\mu' - Q_\mu P_\mu')}{Z} -\partial_\mu \left(\frac{X_\mu}{Z}\right) \right) \right)
\nonumber
\end{eqnarray}
The vanishing of the $u^\lambda$ term follows from the fact that $Z=P_\lambda Q_\mu-P_\mu Q_\lambda$. We deduce that for a particle on an equatorial geodesic ($u^\mu=0$) which has $\rmd\chi/\rmd t = u^\chi/u^t = \omega_c$ initially, then $\rmd\chi/\rmd t = \omega_c(\lambda(t))$ along the entire orbit. This confirms our intuition based on the Newtonian analogue. We may also ask if there are any other $u^\mu=0$ charged particle orbits with this property. The modification to the $\mu$ equation of motion in the presence of charge is a term $(q/G m) F_{\mu \alpha}u^\alpha$. This term reduces to
\begin{equation}
F_{\mu\alpha}u^\alpha = \left(P_\lambda u^t - Q_\lambda u^\chi\right) \left( \partial_\mu \left( \frac{X_\mu}{Z} \right) + \frac{X_\lambda}{Z^2} \left(Q_\mu P_\mu'-P_\mu Q_\mu'\right) \right).
\end{equation}
Thus the equation for $\dot{\mu}$ is unchanged for charged particles when $\omega = \omega_c$. Since no non-equatorial $\mu=const.$ timelike geodesics with $\rmd\chi/\rmd t = \omega_c$ exist, we conclude that $\mu = const.$ timelike orbits traversed with the critical angular velocity occur in the equatorial plane only.

As a final remark, it is clear that at any radius, $\lambda =\lambda_0$, in the equatorial plane, circular charged particle orbits exist. There is a critical charge to mass ratio, $q/G m$, for which the angular velocity of such a particle is equal to the critical angular velocity. This charge to mass ratio is
\begin{equation}
\frac{q}{G m} = \frac{Z^2 \, \partial_\lambda \left( \sqrt{\Delta_\lambda/Z}\right)}{X_\mu\left(Q_\lambda P_\lambda' - P_\lambda Q_\lambda' \right) - Z^2 \, \partial_\lambda \left( X_\lambda/Z\right)} .
\end{equation}
It is understood that all quantities are to be evaluated in the equatorial plane $\mu = 0$ and for $\lambda =\lambda_0$. For any solution that is asymptotically Kerr-Newman, this ratio approaches $e/GM$ as $\lambda \rightarrow \infty$, i.e., the charge to mass ratio of the Kerr-Newman spacetime.

\section{Properties of the solutions}
\label{props}
\subsection{Asymptotic forms}
The metric \erf{cartermet} and electromagnetic field \erf{carterEMF} are of the most general form that allows separability of the Hamilton-Jacobi and Schr\"{o}dinger equations, but they do not necessarily satisfy the vaccum Einstein-Maxwell equations. Requiring the spacetime to satisfy the source-free Einstein-Maxwell equations puts constraints on the various functions of $\mu$ and $\lambda$. Carter finds four allowed families of solutions, for which the electromagnetic field tensor, $A_{\gamma}\,\rmd x^{\gamma}$, takes the forms\footnote[1]{We have interchanged the Killing vectors as compared to Carter's paper \cite{carter68}, so that the results are consistent with interpreting $t$ as the timelike Killing vector.}
\begin{eqnarray}
\left[{\it A}\right]: && e\left\{\frac{\lambda\mu(\mu\cos\alpha+\lambda\sin\alpha)}{\lambda^2 + \mu^2}\rmd \chi + \frac{\lambda\cos\alpha-\mu\sin\alpha}{\lambda^2 + \mu^2} \rmd t \right\} \nonumber \\
\left[\tilde{B}(+)\right]: & &e\left\{\frac{\mu(2l\lambda\cos\alpha+(\lambda^2+l^2)\sin\alpha)}{\lambda^2+l^2}\rmd \chi + \frac{\lambda\cos\alpha}{\lambda^2 +l^2} \rmd t \right\}   \nonumber\\ 
\left[\tilde{B}(-)\right]: & & e\left\{\frac{\lambda((\mu^2+k^2)\cos\alpha+2k\mu\sin\alpha)}{\mu^2+k^2}\rmd \chi+ \frac{\mu\sin\alpha}{\mu^2 +k^2} \rmd t \right\}  \nonumber\\ 
\left[{\it D}\right]: & & e\left\{\lambda\cos\alpha\rmd \chi - \mu\sin\alpha \rmd t \right\} 
\end{eqnarray}
The expression in square brackets at the left-hand side of each field is the label given to each class by Carter \cite{carter68}. Spatial infinity corresponds to the limit $\lambda \rightarrow \infty$. In this limit, the azimuthal part, $A_{\chi}$ of the $\tilde{B}(-)$ solution is proportional to $\lambda$, while for the $D$ solution, either $A_{\chi} \propto \lambda$ or $A_t \propto \mu$. In both cases, this asymptotic form corresponds to a constant electromagnetic field at large distances. The same is true for solutions $A$ and $\tilde{B}(+)$ if $\sin\alpha \neq 0$. If $\sin\alpha = 0$, solution $\tilde{B}(+)$ reduces to the charged NUT solution, with metric
\begin{eqnarray}
\rmd s^2 &=& (\lambda^2+l^2)\left\{\frac{\rmd\lambda^2}{\Delta_\lambda} + \frac{\rmd\mu^2}{\Delta_\mu}+\Delta_\mu\rmd\chi^2\right\} - \frac{\Delta_\lambda[\rmd t+2l\mu\rmd\chi]^2}{\lambda^2+l^2} \nonumber \\
{\rm where} && \Delta_{\lambda} = \Lambda \left(\frac{1}{3}\lambda^4+2l^2\lambda^2-l^4\right) + h(\lambda^2-l^2)-2GM\lambda+Ge^2, \qquad \Delta_{\mu} = -h\mu^2+2q\mu+p
\end{eqnarray}
The NUT solution \cite{nut} represents a gravomagnetic monopole, and has a singularity at $\Delta_\mu=0$, although this can be avoided if the time coordinate is taken to be periodic \cite{misner}. 

We deduce that the only source-free solution in Carter's class that does not include an externally imposed magnetic field or a monopole singularity is $A$ with $\sin\alpha=0$. The metric corresponding to $A$ is\footnote[3]{We have replaced $t$ by $t+(p/h)\chi$ for consistency with \erf{KNfncs}--\erf{KNmet}.}
\begin{eqnarray}
\fl\rmd s^2 &=& (\lambda^2 + \mu^2) \left\{ \frac{\rmd \lambda^2}{\Delta_\lambda} + \frac{\rmd \mu^2}{\Delta_{\mu}}\right\} + \frac{\Delta_\mu [\rmd t - (\lambda^2 + p/h)\rmd\chi]^2 - \Delta_\lambda [\rmd t - (p/h - \mu^2) \rmd\chi]^2}{\lambda^2+\mu^2} \nonumber \\
\fl{\rm where} && \Delta_{\lambda} = \frac{1}{3} \Lambda \lambda^4 + h\lambda^2 - 2mG\lambda + p + G\,e^2, \qquad \Delta_{\mu} =   \frac{1}{3} \Lambda \mu^4 - h\mu^2 +2q\mu + p
\label{asymForm}.
\end{eqnarray}
A coordinate transformation of the form $\lambda \rightarrow \alpha \lambda'$, $\mu \rightarrow \alpha \mu'$, $t \rightarrow t'/\alpha$, $\chi \rightarrow \chi'/\alpha^3$, $e \rightarrow \alpha^2 e'$ and $p \rightarrow \alpha^2 p'$, with $\alpha^2 = |h|$ can be used to set $h=\pm 1$. The requirement that the asymptotic region corresponds to $\lambda \rightarrow \infty$ then forces $h=1$ (taking $h < 1$ essentially changes the role of the coordinates $\mu$ and $\lambda$, but we have chosen to interpret $\lambda$ as the ``radial'' coordinate). The parameter $q$ is a NUT parameter, and to ensure no gravomagnetic monopole in the spacetime we need $q=0$. Finally, the requirement that $\Delta_{\mu} > 0$ means $p/h >0$, so $p=a^2$. Equation \erf{asymForm} is then the Kerr-Newman solution with a cosmological constant.

If we have a non-vaccum spacetime of Carter's type, assume the matter and charge are confined to a finite volume, that there is no externally imposed electromagnetic field and no monopole singularity, then the metric must take the above form outside all the matter in the spacetime. Assuming zero cosmological constant (or that we are in the asymptotically flat regime of the spacetime, but not at cosmological distances), the metric must become \erf{KNmet}. Such solutions thus have the same gyromagnetic ratio as the Kerr-Newman solution, which is known to be the same as for the Dirac electron \cite{carter68b,newman02}. 

\subsection{Asymptotically Kerr-Newman solutions}
If beyond some surface $S: \lambda = \lambda_0$, the metric is Kerr-Newman, what freedom is left in the metric within $S$ if we require it to be of Carter separable form everywhere? Since the functions $P_{\mu}$, $Q_{\mu}$, $\Delta_{\mu}$ and $X_{\mu}$ are functions of $\mu$ alone, they must be the Kerr-Newman functions, $P_{\mu}=1$, $Q_{\mu}=a^2-\mu^2$, $\Delta_\mu = a^2-\mu^2$ and $X_{\mu}=0$ inside $S$ just as they are outside $S$. However, with the exception of $P_{\lambda}$\footnote[1]{Since $Q_{\mu}=a^2-\mu^2$, the requirement that $Z$ must be a function of $\mu$ plus a function of $\lambda$ is satisfied if and only if $P_{\lambda}$ is constant throughout the spacetime.}, the functions of $\lambda$ are free inside $S$ provided that they match the external values at $S$ and the stress tensor generated by them must obey the energy conditions. Of course, if there are discontinuities in their gradients on $S$ then we must ensure that the resultant surface distributions of charge, mass, stress etc. also obey the energy conditions and we should ensure the correct boundary conditions at $\lambda = 0$. However, if it is possible to satisfy all these conditions, we can generate spacetimes that represent interior solutions for the Kerr-Newman spacetime, and which are also globally Schr\"{o}dinger and Hamilton-Jacobi separable. Our attempts to find such solutions have so far failed. In particular, we have shown that no solutions which have the Kerr-Newman form for $Q_{\lambda}$, $Q_{\lambda}=-(\lambda^2+a^2)$, exist that are devoid of a naked ring singularity at $\lambda=\mu=0$ and simultaneously obey the energy conditions everywhere.

\section{Summary}
We have proven a theorem about the electromagnetic fields in Carter separable spacetimes, namely that for every spheroidal surface, $\lambda = const.$, there exists a critical rotation rate, $\omega$, such that the electric and magnetic fields observed in that rotating frame are parallel on that spheroid, and perpendicular to that spheroid. This angular velocity is the angular velocity of the Carter tetrad frame, i.e., the frame in which photons move with purely radial velocity. This generalises a result previously derived for the `Magic' field of the Kerr-Newman black hole which is one member of this class. The result in that case has proven useful for studies of electromagnetic fields in the vicinity of black holes \cite{znaj77}. Our generalised result implies that the separability condition on the electromagnetic field for this family of spacetimes can be interpreted as the requirement that the electric and magnetic fields measured along the null geodesic congruences are parallel and directed along the geodesics in the sense of being in the `radial' direction parallel to $\nabla\lambda$. One consequence of this theorem is that a spheroidal shell of charge, rotating at the critical angular velocity, will experience a force purely normal to its surface.

The Kerr-Newman solution has attracted some interest since it has the same gyromagnetic ratio as the Dirac electron \cite{newman02,newman05,dlb03,dlb04a,dlb04b,carter68b}. As described in Section~\ref{props}, all the Carter separable solutions that have a finite source, have no externally imposed electromagnetic field and no gravomagnetic monopole are asymptotically Kerr-Newman (in fact, identically Kerr-Newman outside the source). This family of spacetimes therefore all have the same gyromagnetic ratio as the Dirac electron. They could thus be used as globally separable spacetime models for further studies of the classical origin of the Dirac gyromagnetic ratio, or as globally separable interior solutions for the Kerr-Newman metric. The theorem presented here demonstrates another generic feature of this class of solutions which we hope will be beneficial in future work.

\acknowledgments We thank Brandon Carter for consultations and for 
stimulating us to 
generalise our flat-space result to all separable space-times. JG's work was supported by St.Catharine's College, Cambridge.



\begin{thebibliography}{10}
\bibitem{ctf} Landau L and Lifshitz E, 1999, {\it Classical Theory of Fields} fourth edition (Oxford: Pergamon Press)
\bibitem{gairthesis} Gair J R, 2002, {\it Generalised Tolman-Bondi Cosmologies and Kerr Metric Atoms}, thesis, Cambridge University
\bibitem{dlb00} Lynden-Bell D, 2000, {\it Carter separable electromagnetic fields}, \MNRAS {\bf 312}, 301
\bibitem{znaj77} Znajek R L, 1977, {\it Black Hole electrodynamics and the Carter tetrad}, \MNRAS {\bf 179}, 457
\bibitem{newman02} Newman E T, 2002, {\it Classical geometric origin of magnetic moments, spin-angular momentum and the Dirac gyromagnetic ratio}, \PRD {\bf 65}, 104005
\bibitem{newman05} Newman E T, 2005, {\it Electromagnetic dipole radiation fields, shear-free congruences and complex centre of charge world lines}, \CQG {\bf 22}, 4667
\bibitem{dlb03} Lynden-Bell D, 2003, {\it A Magic Electromagnetic Field} in {\it Stellar Astrophysical Fluid Dynamics} Eds. Thompson M J and Christiansen-Dalsgaard J (Cambridge: University Press) 
\bibitem{dlb04a} Lynden-Bell D, 2004, {\it Relativistically spinning charged sphere}, \PRD {\bf 70}, 4021L
\bibitem{dlb04b} Lynden-Bell D, 2004, {\it Electromagnetic magic: The relativistically rotating disk}, \PRD {\bf 70}, 5017L
\bibitem{carter68} Carter B, 1968, {\it Hamilton-Jacobi and Schr\"{o}dinger separable solutions of Einstein's equations}, {\it Comm. Math. Phys.} {\bf 10}, 280
\bibitem{carter73} Carter B, 1973, in {\it Les Astres Ocules} (New York: Gordon and Breach) p57-214
\bibitem{nut} Newman E, Tamburino L and Unti T, 1963, {\it Empty space generalization of the Schwarzschild metric}, {\it J. Math. Phys.} {\bf 4} 915
\bibitem{misner} Misner C W, 1967, {\it Relativity Theory and Astrophysics} Vol. 1 ed. J Ehlers (Providence, Rhode Island: Amer. Math. Soc.)
\bibitem{carter68b} Carter B, 1968, {\it Global Structure of the Kerr family of gravitational fields}, \PR {\bf 174} 1559
\end{thebibliography}
\end{document}